\documentclass[a4paper,11pt]{article}
\pdfoutput=1 

\usepackage{jcappub} 

\usepackage[T1]{fontenc} 
\usepackage{subfigure}

\newcommand{\ba}{\begin{eqnarray}}
\newcommand{\ea}{\end{eqnarray}}

\newcommand{\B}{{\cal{B}}}

\newcommand{\DD}{{\cal {D}}}

\newcommand{\W}{{\cal {W}}}
\newcommand{\bbq}{\begin{quote}}
\newcommand{\eeq}{\end{quote}}

\newcommand{\tbb}{t_{\textrm{\tiny{bb}}}}

\newcommand{\RR}{{}^{(3)}{\cal{R}}}

\newcommand{\GG}{{\cal{G}}}
\newcommand{\JJ}{{\cal{J}}}

\newcommand{\HH}{{\cal{H}}}
\newcommand{\KK}{{\cal{K}}}
\newcommand{\PP}{{\cal{P}}}
\newcommand{\QQ}{{\cal{Q}}}

\newcommand{\Drho}{\Delta^{(\rho)}}

\newcommand{\Huno}{\textrm{\bf{H}}^{(1)}}
\newcommand{\Hdos}{\textrm{\bf{H}}^{(2)}}
\newcommand{\Htres}{\textrm{\bf{H}}^{(3)}}
\newcommand{\Hii}{\textrm{\bf{H}}^{(A)}}

\newcommand{\bW}{{\rm{\bf W}}}

\newcommand{\DDa}{{\textrm{\bf{D}}}^{(A)}}

\newcommand{\DDrho}{{\textrm{\bf{D}}}^{(\rho)}}

\newcommand{\DDKK}{{\textrm{\bf{D}}}^{(\KK)}}

\newcommand{\DDh}{{\textrm{\bf{D}}}^{(\HH)}}

\newcommand{\muqls}{[\mu_q]_{\textrm{\tiny{LS}}}}
\newcommand{\kaqls}{[\kappa_q]_{\textrm{\tiny{LS}}}}
\newcommand{\tls}{t_{\textrm{\tiny{LS}}}}
\newcommand{\Hls}{\bar H_{\textrm{\tiny{LS}}}}
\newcommand{\als}{a_{\textrm{\tiny{LS}}}}
\newcommand{\Als}{A_{\textrm{\tiny{LS}}}}
\newcommand{\barAls}{\bar{A}_{\textrm{\tiny{LS}}}}
\newcommand{\Gls}{\Gamma_{\textrm{\tiny{LS}}}}
\newcommand{\DDals}{{\textrm{\bf{D}}}^{(A)}_{\textrm{\tiny{LS}}}}
\newcommand{\Drhols}{\Delta^{(\rho)}_{\textrm{\tiny{LS}}}}
\newcommand{\ini}{{\textrm{{ini}}}}
\newcommand{\vpcmb}{v_{{\textrm{\tiny{pec}}}}^{{\textrm{\tiny{cmb}}}}}
\newcommand{\vpvoid}{v_{{\textrm{\tiny{pec}}}}^{{\textrm{\tiny{void}}}}}
\newcommand{\Aav}{\langle A\rangle}
\newcommand{\rhoav}{\langle \rho\rangle}
\newcommand{\HHav}{\langle \HH\rangle}

\newcommand{\KKav}{\langle \KK\rangle}

\newcommand{\Drhonl}{\Delta_{\textrm{\tiny{NL}}}^{(\rho)}}

\newcommand{\DDanl}{{\textrm{\bf{D}}}_{\textrm{\tiny{NL}}}^{(A)}}
\newcommand{\rhobaras}{\bar\rho_{\textrm{\tiny{as}}}}
\newcommand{\HHbaras}{\bar\HH_{\textrm{\tiny{as}}}}
\newcommand{\KKbaras}{\bar\KK_{\textrm{\tiny{as}}}}
\newcommand{\Abaras}{\bar A_{\textrm{\tiny{as}}}}
\newcommand{\DDaas}{{\textrm{\bf{D}}}_{\textrm{\tiny{as}}}^{(A)}}
\newcommand{\Drhoas}{\Delta_{\textrm{\tiny{as}}}^{(\rho)}}
\newcommand{\Dhas}{{\textrm{\bf{D}}}_{\textrm{\tiny{as}}}^{(\HH)}}
\newcommand{\DKKas}{{\textrm{\bf{D}}}_{\textrm{\tiny{as}}}^{(\KK)}}

\title{\boldmath Coarse--grained description of cosmic structure from Szekeres models}

\author[a]{Roberto A. Sussman,}
\author[b]{I. Delgado Gaspar,}
\author[c]{and Juan Carlos Hidalgo}

\affiliation[a]{Instituto de Ciencias Nucleares, Universidad Nacional Aut\'onoma de M\'exico,
A. P. 70--543, 04510 M\'exico D. F., M\'exico.}
\affiliation[b]{Instituto de Investigaci\'on en Ciencias B\'asicas y Aplicadas, Universidad Aut\'onoma del Estado de Morelos, Av Universidad 1002, 62210, Cuernavaca, Morelos, M\'exico.}
\affiliation[c]{Instituto de Ciencias F\'{\i}sicas, Universidad Nacional Aut\'onoma de M\'exico, A.P. 48--3, 62251 Cuernavaca, Morelos, M\'exico}

\emailAdd{sussman@nucleares.unam.mx}
\emailAdd{ismael.delgadog@uaem.edu.mx}
\emailAdd{hidalgo@fis.unam.mx}

\abstract{We show that the full dynamical freedom of the well known Szekeres models allows for the description of elaborated 3--dimensional networks of cold dark matter structures (over--densities and/or density voids) undergoing ``pancake'' collapse. By reducing Einstein's field equations to a set of evolution equations, which themselves reduce in the linear limit to evolution equations for linear perturbations, we determine the dynamics of such structures, with the spatial comoving location of each structure uniquely specified by standard early Universe initial conditions. By means of a representative example we examine in detail the density contrast, the Hubble flow and peculiar velocities of structures that evolved, from linear initial data at the last scattering surface, to fully non--linear 10--20 Mpc scale configurations today. To motivate further research, we provide a qualitative discussion on the connection of Szekeres models with linear perturbations and the pancake collapse of the Zeldovich approximation. This type of structure modelling provides a coarse grained -- but fully relativistic non--linear and non--perturbative -- description of evolving large scale cosmic structures before their virialisation, and as such it has an enormous potential for applications in cosmological research.}

\begin{document}
\maketitle
\flushbottom

\section{Introduction}

Considering numerical solutions of Einstein's equations applied to Cosmology is evidently an urgent task, as it is widely assumed to be practically impossible to model minimally realistic cosmic structures (even at a coarse grained level) by means of exact solutions of Einstein's equations. Since cosmological application of numerical General Relativity is still in its early stage of development \cite{Adamek:2014xba,Adamek:2013wja,Fidler:2015npa}, most cosmological applications that require a non--perturbative relativistic approach  still rely on the highly idealised class of spherically symmetric Lema\^\i tre--Tolman--Bondi (LTB) dust models \cite{kras2, BKHC2009}, which can only describe the evolution of a single spherical dust structure embedded in an FLRW background. 

The non--spherical Szekeres dust models (see details of their classification in \cite{kras2, BKHC2009}) are a well known generalisation of LTB models. Although it is wholly unreasonable to expect of these models (themselves an exact solution Einstein's equations) to provide the level of ``realism'' expected from the (yet to develop) numerical solutions, Szekeres models are still useful and have been applied to address various cosmological and observational issues  \cite{Bsz1,Bsz2,IRGW2008,Bole2009-cmb,KrBo2010,Bole2010-sn,BoCe2010,NIT2011,MPT,PMT,BoSu2011,sussbol,WH2012,Buckley,Vrba,kokhan,mock}. However, in practically all this literature the authors consider models that only describe the evolution of two structures (an over--density next to a density void) in simple axial dipolar arrays. In a recent article \cite{nuevo} we showed that the full dynamical freedom of the models allows for the description of far more general configurations, namely: elaborated networks of multiple evolving cosmic structures (over--densities and density voids defined by 3--dimensional maxima and minima of the density), whose spatial (radial and angular) location at all $t$ can be {\it a priori} specified by suitable initial conditions. However, \cite{nuevo} was essentially a theoretical study concerned with the existence conditions of the extrema (maxima, minima and saddle points) of all Szekeres scalars (not just the density) of generic models (ever expading, collapsing, with zero and nonzero $\Lambda$). While various  relevant technical issues, such as avoidance of shell crossings, were extensively discussed, this study only provided a simple qualitative example (see its figures 8 and 9) of the density contrast of a multi--structure Szekeres configuration at an initial time, without studying its actual evolution and without illustrating the shape of other scalars (for example the Hubble scalar). 

In the present article we aim at extending and enhancing the work in \cite{nuevo}, specifically by undertaking the following tasks: (i)
{\bf Modelling more general structure networks.} The structures in \cite{nuevo} only admitted an over--density or a density void in each radial shell zone. More general networks of structures are now obtained, admitting an arbitrary number of structures in assorted angular locations in each radial zone. With this improvement we can now attempt to implement a coarse grained modelling of large scale cosmography, either from observed and/or reconstructed studies \cite{cosmography1,cosmography2} or from numerical simulations \cite{N-body1}. (ii) {\bf Realistic evolution.} Assuming a $\Lambda$CDM background consistent with observations, we examine the numerical evolution of the above mentioned improved networks, from linear perturbations at the last scattering surface into an $\sim$\,80 Mpc region containing 10--20 Mpc sized structures in a non--linear regime at present cosmic time.
(iii) {\bf Expansion, collapse and peculiar velocities.} We obtain and depict the present day anisotropic and inhomogeneous Hubble flow that shows the structures undergoing ``pancake'' collapse at rates appropriate to their length scales. The peculiar velocities of the structures are examined, providing a qualitative comparison with velocities reported in the existing literature. (iv) {\bf Theoretical issues.} We show that the pancake collapsing Szekeres over--densities provide an exact relativistic analogue of the Newtonian  Zeldovich Approximation. We also comment on the correspondence with linear perturbations of dust sources (in the isochronous gauge) and on the  back--reaction issue.     

The section by section contents are as follows. In section 2 we derive a  set of evolution equations (equivalent to Einstein's field equation) that fully determine the dynamics of Szekeres models. A procedure to construct Szekeres configurations describing elaborated networks of structures (over--densities and density voids) is presented in section 3. In section 4 we build up a representative numerical example consisting of a central spheroidal void surrounded by multiple over--densities undergoing pancake collapse. In section 5 we examine in detail (i) the density contrast, the Hubble scalar and eigenvalues of the expansion tensor and radial peculiar velocities associated with the example of section 4. The connection with the Zeldovich approximation and dust linear perturbations are discussed in section 6. Conclusions and guidelines for further research and applications are stated in section 7.    

\section{The dynamics of Szekeres models}\label{dynamics}

Quasi--spherical Szekeres models of class I in ``stereographic'' spherical coordinates $(r,\theta,\phi)$ \cite{kras2,Bsz1,Bsz2} are described by the following non--diagonal metric
\footnote{A simpler diagonal metric (with several variations) is used in most of the existing literature. The transformations relating (\ref{g1})--(\ref{g3}) to this metric are given in Appendix A of \cite{nuevo}.}
\ba 
 g_{tt}&=& -1,\quad g_{rr} =  a^2\Bigg\{ \frac{(\Gamma-\bW)^2}{1-[\KK_{q}]_\ini r^2} 
+\frac{\sin^4\theta}{(1+\cos\theta)^2}\left[\W^2-2\frac{1+\cos\theta}{\sin^2\theta}\,Z\,\bW\right]\Bigg\}, \label{g1}\\
 g_{r\theta}&=&\frac{a^2\,r\,\sin\theta}{1+\cos\theta}\left(\bW-Z\right),\; g_{r\phi}=-\frac{a^2\,r\,\sin^2\theta}{1+\cos\theta}\,\bW_{,\phi},\label{g2}
\\
 g_{\theta\theta} &=& a^2 r^2,\quad g_{\phi\phi} = a^2 r^2 \sin^2\theta ,\label{g3}
\ea
where $a=a(t,r),\,\Gamma=1+ra'/a$, with $a'=\partial a/\partial r$,\,\,$[\KK_{q}]_\ini=[\KK_{q}]_\ini(r)$ (see (\ref{qscals})), and the Szekeres dipole $\bW$ is given by
\begin{equation}
\bW = -X\sin \theta \cos \phi -Y \sin \theta \sin \phi-Z \cos \theta,\label{dipole}
\end{equation}
where $X=X(r), \, Y=Y(r), \, Z=Z(r)$ are the dipole free functions and $\W=\sqrt{X^2+Y^2+Z^2}$ is the dipole magnitude. It is straightforward to see from (\ref{g1})--(\ref{g3}) that the surfaces of constant $t$ and $r$ are 2--spheres (with surface area $4\pi a^2 r^2$) that are non--concentric about the origin worldline \cite{kras2,nuevo}. By setting $X=Y=Z=0\,\,\Rightarrow\,\,\bW=0$ we obtain a generic LTB ``seed model'' as the unique spherically symmetric sub--case. 

The models are fully characterised by their covariant fluid flow scalars: the density $\rho$, the Hubble scalar $\HH=\Theta/3$ and the spatial curvature $\KK=(1/6)\RR$ (with $\Theta=\nabla_au^a$ and $\RR$ the Ricci scalar of the hypersurfaces of constant $t$). Considering a nonzero cosmological constant to accommodate a $\Lambda$CDM background, the dynamics of the models becomes fully determined by the numerical solutions of the following evolution equations derived in \cite{sussbol}:
\footnote{In practically all the existing literature the dynamics of Szekeres models is studied in terms of the integral solutions (analytic or numerical) of the Friedman--like equation (\ref{constraints1}). See for example the comprehensive work in \cite{WH2012}, which only considered models with $\Lambda=0$. We believe that the evolution equations (\ref{FFq1})--(\ref{FFq6}) provide a much more efficient framework for numerical work, specially for models with $\Lambda>0$.}
\ba  \dot\rho_q &=& -3 \rho_q\,\HH_q,\label{FFq1}\\
 \dot \HH_q &=& -\HH_q^2-\frac{4\pi}{3}\rho_q+\frac{8\pi}{3}\Lambda, \label{FFq2}\\
 \dot\Delta^{(\rho)} &=& -3(1+\Drho)\,\DDh\label{FFq3}\\
 \dot {\textrm{\bf{D}}}^{(\HH)} &=&  \left(-2\HH_q+3\DDh\right)\DDh-\frac{4\pi}{3}\rho_q\Drho,\label{FFq4}\\
 \dot a &=& a\,\HH_q, \label{FFq5}\\
 \dot\GG &=& 3\,\GG\,\DDh,\qquad \GG=\frac{\Gamma-\bW}{1-\bW} , \label{FFq6}\ea
subject to the algebraic constraints:
\ba
\HH_q^2 &=&\frac{8\pi}{3}\left[\rho_q+\Lambda\right]-\KK_q,\label{constraints1}\\
2\HH_q\DDh &=& \frac{8\pi}{3}\DDrho-\DDKK,\label{constraints2}\ea
where the ``q--scalars'' $A_q$ and their exact fluctuations $\DDa$ \cite{sussbol} (which determine the standard covariant scalars) are given by
\begin{eqnarray} \rho_q &=& \frac{[\rho_{q}]_\ini}{a^3},\quad \KK_q=\frac{[\KK_{q}]_\ini}{a^2},\quad \HH_q=\frac{\dot a}{a},\label{qscals}\\
\DDa &=& A-A_q=\frac{r\,A'_q}{3(\Gamma-\bW)},\quad A=\rho,\,\HH,\,\KK,\label{DDa}\\
\Drho &=& \frac{\Drho}{\rho_q}=\frac{\rho-\rho_q}{\rho_q}=\frac{r\,\rho'_q/\rho_q}{3(\Gamma-\bW)},\label{Drho}
 \end{eqnarray}      
with the subindex ${}_\ini$ denoting henceforth evaluation at an arbitrary time slice $t=t_\ini$.  

Besides the cosmological constant $\Lambda$, the initial conditions to integrate the system (\ref{FFq1})--(\ref{FFq4}) are the following five free parameters: the two ``radial'' initial functions common to LTB models, $[\rho_{q}]_\ini,\,[\KK_{q}]_\ini$,  and the ``angular dipole'' initial functions $X,\,Y,\,Z$. Initial values of $\HH_q,\,\Drho,\,\DDh,\,\DDKK$ follow from (\ref{constraints1})--(\ref{constraints2}), the radial coordinate is chosen so that $a_\ini=\Gamma_\ini=1$, while the Big Bang time and its gradient follow from the choice of $[\rho_{q}]_\ini,\,[\KK_{q}]_\ini$ (see \cite{sussbol,nuevo}).  

The q--scalars and their fluctuations are coordinate independent objects that are directly related to curvature and kinematic scalars \cite{sussbol,nuevo}. As we show in section 6.2, they reduce in the linear limit to standard variables of cosmological dust perturbations in the synchronous gauge (see the LTB case in \cite{perts}).

\section{Networks of over--densities and density voids}\label{networks}

Over--densities and density voids can be defined as regions surrounding the spatial maxima and minima of the matter density. The coordinate location of the density extrema (as that of all other scalars $A$) follows from the condition $A'=A_{,\theta}=A_{,\phi}=0$, whose solutions are, at each constant $t$,
\begin{equation} r=r_{e\pm},\quad \theta=\theta_\pm(r_{e\pm}),\quad\phi=\phi_\pm(r_{e\pm})\end{equation}
where $\theta_\pm(r)$ and $\phi_\pm(r)$ follow from the solutions of the subset $A_{,\theta}=A_{,\phi}=0$ (the ``angular extrema'' \cite{nuevo}), 
\ba  
 \phi_{-} &=& \arctan \left(\frac{Y}{X}\right),\quad \phi_{+} = \pi+\phi_{-}, \label{sol1}\\
\theta_{-}&=& \arccos\left(\frac{Z}{\sqrt{X^2+Y^2+Z^2}}\right), \quad \theta_{+}= \pi-\theta_{-},\label{sol2}
 \ea
which for every fixed $r$ defines a precise angular direction and also, for varying $r$, the two ``curves of angular extrema''  $\B_{\pm}(r )=[r,\,\theta_{\pm}(r ),\,\phi_{\pm}( r)]$ in all time slices. To find the radial location $r=r_{e\pm}$ of the extrema we need to solve the ``radial'' conditions $A'_\pm(t,r)=0$, where the subindex ${}_\pm$ denotes evaluation along the curves $\B_{\pm}(r )$ (notice that $\bW_\pm=\pm\W$).  

For each solution of $A'_\pm=0$ at arbitrary $t$ there will be an extremum of the scalar $A$ at angular coordinates (\ref{sol1})--(\ref{sol2}). As shown in \cite{nuevo}, a sufficient condition
 for the existence at all $t$ (pending shell crossings) of an arbitrary number of such solutions (and thus an arbitrary number of spatial extrema of all Szekeres scalars) is furnished by assuming compatibility with Periodic Local Homogeneity (PLH), defined by the vanishing for all $t$ of the shear ($\sigma^a_b$) and electric Weyl ($E^a_b$) tensors  along a sequence of comoving 2--spheres (comoving homogeneity spheres) \cite{nuevo}. Since (\ref{constraints2}) and (\ref{DDa}) are preserved by the time evolution, models compatible with PLH are specified by the following initial conditions:   
\footnote{It is important to emphasise that PLH is a sufficient (but not necessary) condition for the existence of spatial maxima and minima of Szekeres scalars. These extrema can also arise without assuming PLH by means of ``simulated shell crossings'' induced by the dipole parameters $X,\,Y,\,Z$ for arbitrary choices of initial value functions $[\rho_{q}]_\ini,\,[\KK_{q}]_\ini,\,[\HH_{q}]_\ini$. See detail in \cite{nuevo}.}
\ba  
&{}&\Rightarrow\quad \DDa_\ini(r_*^i,\theta,\phi)=0\quad \Rightarrow\quad [A'_{q}]_\ini(r_*^i)=0 ,\nonumber\\
&{}&\Rightarrow\quad  A_\ini(r_*^i,\theta,\phi)= [A_{q}]_\ini(r_*^i), \label{PLH0}  
\ea 
which imply that $[\sigma^a_b]_*=[E^a_b]_*=0$ holds for all $t$, with the subindex ${}_*$ denoting evaluation at the sequence of $n$ nonzero values of the radial coordinate $r_*^i,\,i=1,..,n$ that mark the comoving homogeneity spheres.  

Depending on the number of values $r_*^i$ in the comoving shells defined by the intervals $\Delta_*^i=r_*^{i-1}<r<r_*^i$, initial conditions (\ref{PLH0}) lead to $2n+1$ spatial extrema distributed
 in the $\Delta_*^i$
 as follows (see comprehensive discussion in \cite{nuevo}):
\begin{itemize}
\item If regularity conditions hold \cite{kras2,nuevo} the origin $r=0$ is always a spatial maximum or a spatial minimum (depending on the sign of $A''$ at $r=0$).     
\item There are $n$ maxima or minima along $\B_{+}$ (this depends on the profile of $A_\pm$ 
and the sign of $\DDa$, 
see figure 1), while the $n$ extrema along $\B_{-}(r)$ are necessarily spatial saddles.
\end{itemize}  
However, as we show in this article, the models admit more general configurations with an arbitrary number of spatial maxima or minima at each radial interval $\Delta_*^i$. Further restrictions on the functions $[A_{q}]_\ini$ satisfying conditions (\ref{PLH0}) may be necessary to 
ensure the
absence of shell crossings and the preservation of the initial concavity for all the evolution (see sections IX and X of \cite{nuevo}).

It important to emphasise that the comoving shells marked by the radial interval $\Delta_*^i$ are not ``FLRW regions'', since $\sigma^a_b=E^a_b=0$ only holds in the boundaries $r=r_*^i$ of these shells (the ``comoving homogeneity spheres''). Since Szekeres models can always be matched (along comoving 2--spheres) to regions of dust FLRW models \cite{BoCe2010}, any one of the comoving shells marked by some $\Delta_*^i$ can be replaced by a FLRW shell region matched to contiguous shells (so that $\sigma^a_b=E^a_b=0$ holds in the whole interval $\Delta_*^i$), but this type of matching is too restrictive and thus has not been considered. However, even if we had introduced such FLRW shell regions, the resulting configuration would bear no  resemblance to Swiss Cheese models, as in the latter the FLRW ``cheese'' is not distributed in spherical shells but is surrounding a collection of inhomogeneous dust vacuoles (the ``holes''). Further discussion on these issues is given in the conclusions section.   
\begin{table}
\begin{center}
\begin{tabular}{|c| c| c|}
\hline
\multicolumn{3}{|c|}{$\Delta_*^1 = 0<\chi<\chi_*^1$}
\\
\hline
\multicolumn{3}{|c|}{$\muqls=\QQ_{0,1}(\chi),\qquad \kaqls=\PP_{0,1}(\chi)$}
\\  
\hline
{$\Delta\phi_{1k}$} &{$X$} &{$Y$} 
\\
\hline
{$\psi_{11}<\phi<\psi_{12}$} &{$-\cos\phi_{11}\zeta_{11}f_1$} &{$-\sin\phi_{11}\zeta_{11}f_1$}
\\  
\hline
{$\psi_{12}<\phi<2\pi+\psi_{11}$} &{$-\cos\phi_{12}\zeta_{12}f_1$} &{$-\sin\phi_{12}\zeta_{12}f_1$} 
\\
\hline
\hline
\multicolumn{3}{|c|}{$\Delta_*^2 = \chi_*^1<\chi<\chi_*^2$}
\\
\hline
\multicolumn{3}{|c|}{$\muqls=\QQ_{1,2}(\chi),\qquad \kaqls=\PP_{1,2}(\chi)$}
\\  
\hline
{$\Delta\phi_{2k}$} &{X} &{Y} 
\\
\hline
{$\psi_{21}<\phi<\psi_{22}$} &{$-\cos\phi_{21}\zeta_{21}f_2$} &{$-\sin\phi_{21}\zeta_{21}f_2$} 
\\
\hline
{$\psi_{22}<\phi<2\pi+\psi_{21}$} &{$-\cos\phi_{22}\zeta_{22}f_2$} &{$-\sin\phi_{22}\zeta_{22}f_2$}  
\\
\hline
\hline
\multicolumn{3}{|c|}{$\Delta_*^3 = \chi_*^2<\chi<\chi_*^3$}
\\
\hline
\multicolumn{3}{|c|}{$\muqls=\QQ_{2,3}(\chi),\qquad \kaqls=\PP_{2,3}(\chi)$}
\\  
\hline
{$\Delta\phi_{3k}$} &{X} &{Y} 
\\
\hline
{$\psi_{31}<\phi<\psi_{32}$} &{$-\cos\phi_{31}\zeta_{31}f_3$} &{$-\sin\phi_{31}\zeta_{31}f_3$} 
\\
\hline
{$\psi_{32}<\phi<\psi_{33}$} &{$-\cos\phi_{32}\zeta_{32}f_3$} &{$-\sin\phi_{32}\zeta_{32}f_3$} 
\\
\hline
{$\psi_{33}<\phi<2\pi+\psi_{31}$} &{$-\cos\phi_{33}\zeta_{33}f_3$} &{$-\sin\phi_{33}\zeta_{33}f_3$} 
\\
\hline
\hline
\multicolumn{3}{|c|}{$\Delta_*^4 = \chi_*^3<\chi<\chi_*^4$}
\\
\hline
\multicolumn{3}{|c|}{$\muqls=\QQ_{3,4}(\chi),\qquad \kaqls=\PP_{3,4}(\chi)$}
\\  
\hline
{$\Delta\phi_{4k}$} &{X} &{Y} 
\\
\hline
{$\psi_{41}<\phi<\psi_{42}$} &{$-\cos\phi_{41}\zeta_{41}f_4$} &{$-\sin\phi_{41}\zeta_{41}f_4$} 
\\
\hline
{$\psi_{42}<\phi<\psi_{43}$} &{$-\cos\phi_{42}\zeta_{42}f_3$} &{$-\sin\phi_{42}\zeta_{42}f_4$} 
\\
\hline
{$\psi_{43}<\phi<2\pi+\psi_{41}$} &{$-\cos\phi_{43}\zeta_{43}f_4$} &{$-\sin\phi_{43}\zeta_{43}f_4$} 
\\
\hline
\end{tabular}
\end{center}
\caption{{\bf{Initial conditions at last scattering}}. The table displays the piecewise definition of the free functions $\muqls,\,\kaqls,\,X$ and $Y$ needed to integrate the system (\ref{FFq1})--(\ref{constraints2}) (as we have assumed $Z=0$ and $a=\Gamma=1$ at $t=t_i=\tls$). The fifth order polynomials $\QQ_{j-1,k}$ and $\PP_{j-1,k}$ are defined by conditions (\ref{mukaconds}). The normalised coordinates of the comoving homogeneity spheres are $[\chi_*^1,\chi_*^2,\chi_*^3,\chi_*^4]\,=\,[3.47826, 6.06906, 7.99883, 9.43622]$, with $\chi_*^0=0$. The azimuthal angular location of the maxima are given by $[\phi_{11},\phi_{12},\phi_{21},\phi_{22},\phi_{31},\phi_{32},\phi_{33},\phi_{41},\phi_{42},\phi_{43}]\,=\,[0,5\pi/4,3\pi/4,7\pi/4,0,3\pi/3,4\pi/3,\pi/3,\pi,5\pi/3]$. The boundaries of the azimuthal partitions are $[\psi_{11},\psi_{12},\psi_{21},\psi_{22},\psi_{31},\psi_{32},\psi_{33},\psi_{41}, \psi_{41},\psi_{42},\psi_{43}]\,=\,[1.96466,5.10625,\pi/4,5 \pi/4,2\pi/6,\pi,  5 \pi/3,0,2\pi/3,4\pi/3]$. The amplitude constants are $\zeta_{11}=0.885,\,\zeta_{12}=0.89$,\,$\zeta_{21}=\zeta_{22}=\zeta_{23}=0.823$, $\zeta_{31}=\zeta_{32}=\zeta_{33}=0.844$ and  $\zeta_{41}=\zeta_{42}=\zeta_{43}=0.8592$. }
\label{tabla1}
\end{table}
            
\section{Numerical example of multiple structures.}\label{example}

We illustrate the set-up of a Szekeres model describing multiple evolving structures through a simple idealised  numerical example (more elaborated examples can easily be obtained along these lines). For this purpose, we assume an asymptotic $\Lambda$CDM background characterised by the present day parameters from Planck 2013 \cite{planck}: $\bar\Omega_0^m=0.32$ (includes baryons),\,$\bar\Omega_0^\Lambda=0.68$ and $\bar H_0=68\,\hbox{km/s Mpc}$ (over--bar denotes background $\Lambda$CDM variables). We consider the model evolution (governed by (\ref{FFq1})--(\ref{constraints2})) from linear initial conditions (see Table 1) at the last scattering surface (LS) $t_\ini=\tls\sim 4\times 10^5\,\hbox{ys}$ that also comply with PLH conditions (\ref{PLH0}).      
\begin{figure}
\centering
\includegraphics[scale=0.3]{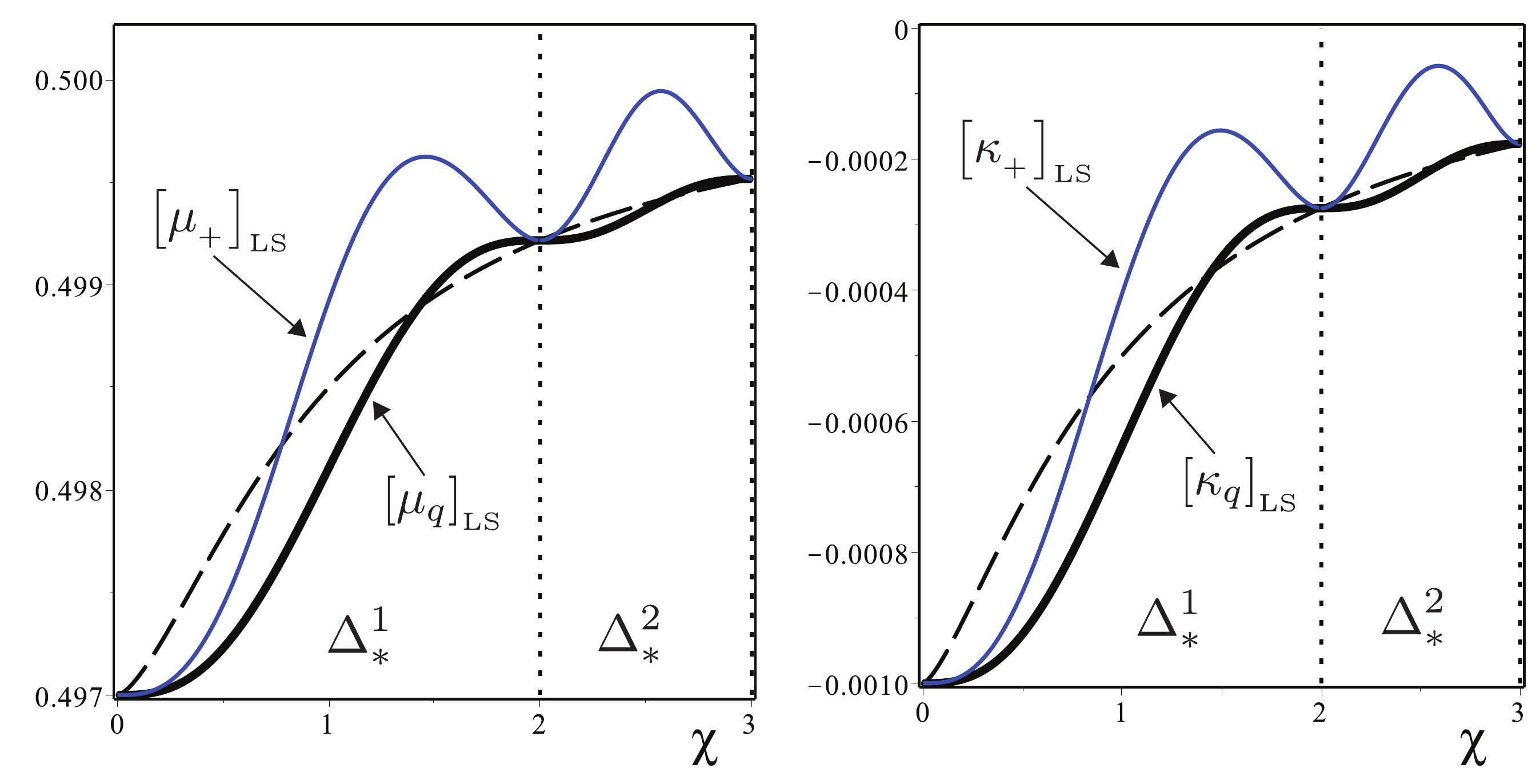}
\caption{{\bf Profiles of the ``radial'' initial value functions $\muqls$ and $\kaqls$.} These functions, defined by (\ref{mukaconds}) in terms of fifth order polynomials at each interval $\Delta_*^i$, are depicted as solid black curves (only two  intervals $\Delta_*^i$ are displayed). The dashed curves are the auxiliary functions $m(\chi)$ and $k(\chi)$ defined in the text and the blue curves are the initial functions $\mu_{\textrm{\tiny{LS}}}$ and $\kappa_{\textrm{\tiny{LS}}}$ evaluated along the curves $\B_{+}(r)$, which are a collection of piecewise continuous line segments (see figures 2, 3 and 4). The maxima of these curves provide the radial coordinate of all maxima at every $\Delta_*^i$.}
\label{Fig1}
\end{figure}

\subsection{Radial location of the spatial density maxima}\label{radial}

We consider the dimensionless initial density and spatial curvature q--scalars (\ref{qscals}) as
\ba \muqls(\chi) =\frac{4\pi [\rho_q]_{\textrm{\tiny{LS}}}}{3 \Hls^2},\quad \kaqls(\chi) =\frac{[\KK_q]_{\textrm{\tiny{LS}}}}{\Hls^2},\label{mukaqls}\ea
where $\Hls\sim 2/(3\tls)$ and $\chi=r/r_s$ with $r_s = 0.0025$ Mpc, which fixes the unit comoving length scale well within the comoving horizon scale at LS. An initial density minimum at $r=0$ follows from the condition $\muqls''(0)>0$. For a sequence of density maxima inside intervals $\Delta_*^i$ along the curve $\B_{+}(r)$ (as in figure 8 of \cite{nuevo}), the profiles of $\muqls$ and $\kaqls<0$ for all $\chi$ must correspond to non--decreasing functions complying with (\ref{PLH0}) and with a $\Lambda$CDM background, whose profiles are depicted by figure 1 (see also panel (a) of figure 4 in \cite{nuevo}). We specify $\muqls$ and $\kaqls$ as piecewise functions defined at each interval $\Delta_*^i$ for a sequence of four intervals $\chi_*^i$ with $\chi_*^0=0$ (see Table 1), where $\QQ_{i-1,i}(\chi)$ and $\PP_{i-1,i}(\chi)$ are fifth order polynomials whose six coefficients are determined (at each $\Delta_*^i$) by two boundary conditions and four conditions to fulfill smoothness (of the metric, the covariant scalars and their first derivatives) at the dimensionless comoving radii $\chi_*^i$: 
\ba \QQ_{i-1,j}(\chi_*^{i-1}) &=& m(\chi_*^{i-1}),\quad \QQ_{i-1,i}(\chi_*^i)=m(\chi_*^i),
\quad \QQ'_{i-1,i}=\QQ''_{i-1,j}=0,\quad \hbox{at}\,\,\chi=\chi_*^{i-1},\chi_*^i,\nonumber\\
\PP_{i-1,i}(\chi_*^{i-1}) &=& k(\chi_*^{i-1}),\quad \PP_{i-1,j}(\chi_*^i)=k(\chi_*^i),
\quad \PP'_{i-1,i}=\PP''_{i-1,i}=0,\quad \hbox{at}\,\,\chi=\chi_*^{i-1},\chi_*^i,\nonumber\\
\label{mukaconds}
\ea
with the auxiliary functions  $m(\chi)=0.5-0.3/(1+\chi^3)$ and $k(\chi)=-0.0014/(1+\chi^{7/5})$ (dashed black curves in figure 1). 
The normalised radial coordinates $\chi_{e+}^i$ of the initial density maxima in each $\Delta_*^i$ are the maxima 
of the curve $[\mu_{+}]_{\textrm{\tiny{LS}}}$ in the left panel of figure 1. To ensure a $\Lambda$CDM background at $\tls$ we set $\muqls=m(\chi)$ and $\kaqls=k(\chi)$ for $\chi>\chi_*^4$, leading to $2\muqls\to \bar\Omega_{\textrm{\tiny{LS}}}^m= 1$ and $\kaqls\to\bar\Omega_{\textrm{\tiny{LS}}}^k=0$ as $r\to\infty$ 
(in our setup $\bar\Omega_{\textrm{\tiny{LS}}}^\Lambda\sim 10^{-9}$). 
The functions $\muqls$ and $\kaqls$ comply with the following desirable properties: (i) they are consistent with the conditions to avoid shell crossings for the time range $\tls<t<t_0$; (ii) they produce a non--simultaneous Big Bang time ($\tbb'\ne 0$), but with negligible differences (of order $\sim 10^3\,\hbox{ys}$)  
in the cosmic age for all observers at $t_0$, and (iii) the concavity of the central void and the density maxima is preserved for all $t>\tls$ (these technical issues are discussed in detail in \cite{nuevo}). 
\begin{figure*}
\centering
\includegraphics[scale=0.59]{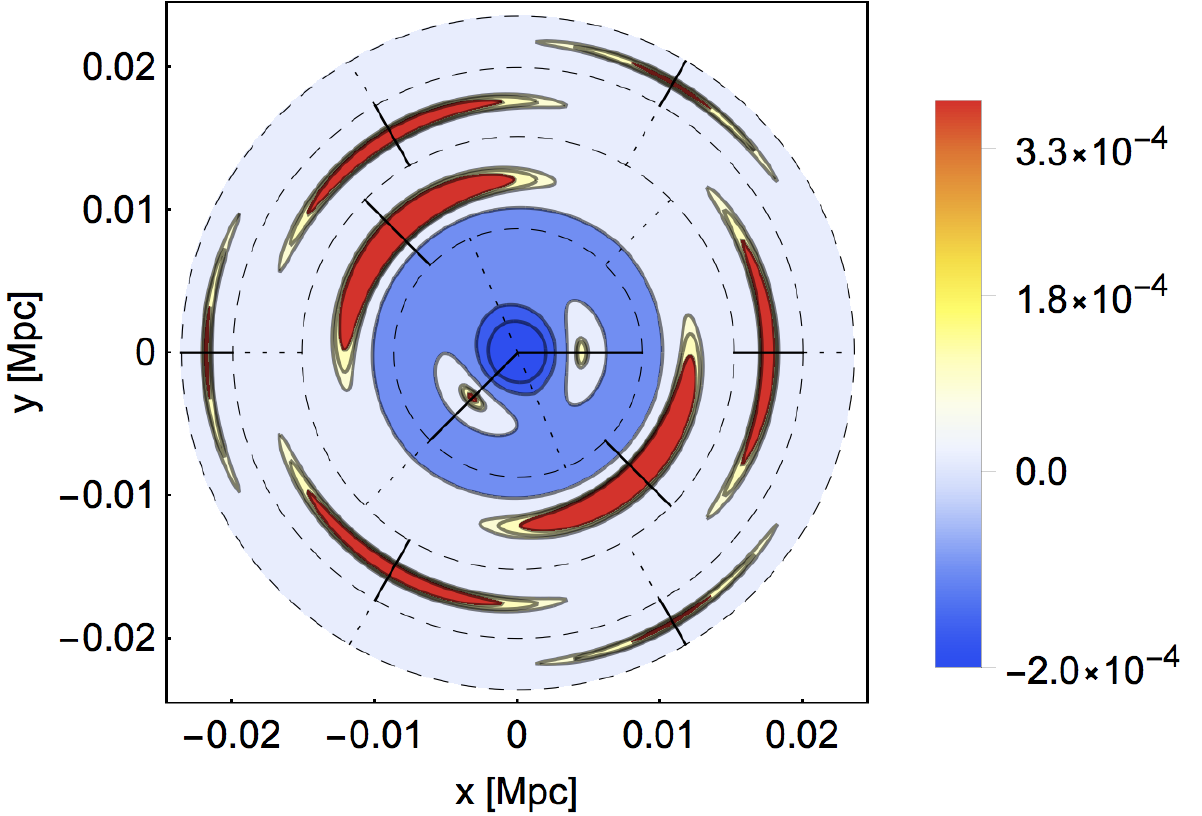}
\quad
\includegraphics[scale=0.55]{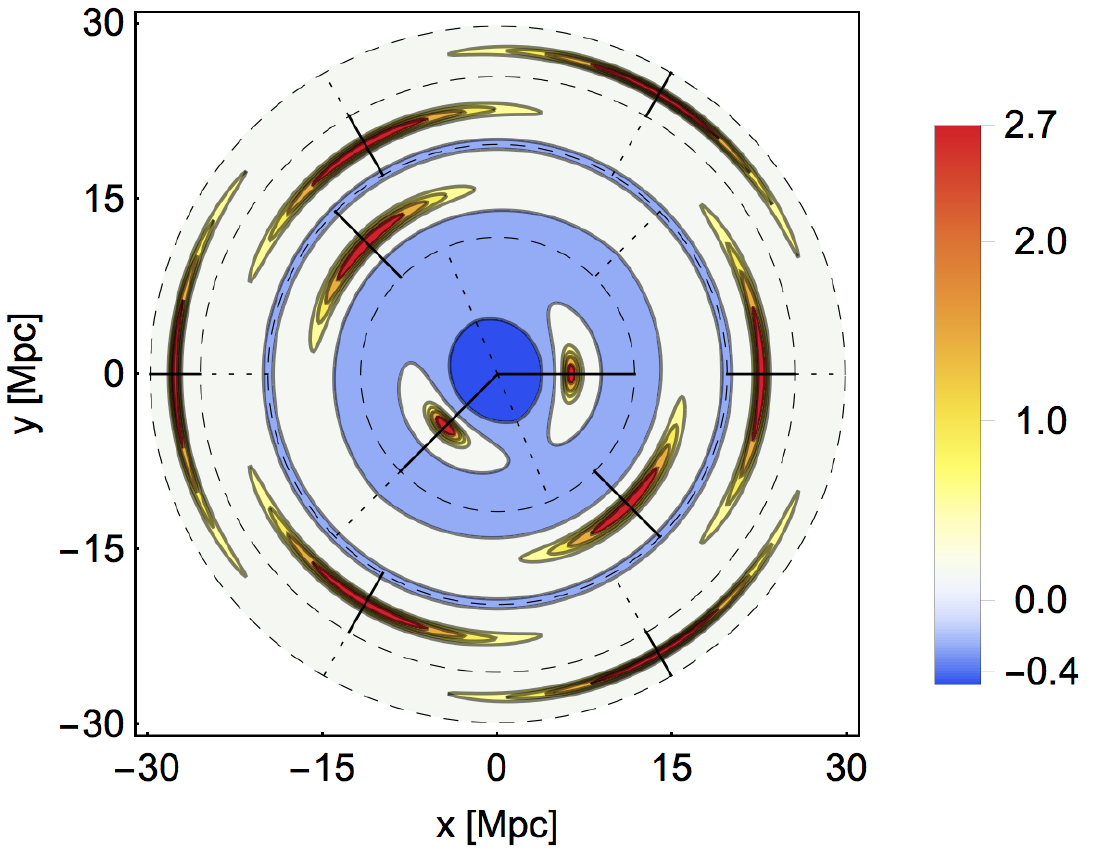}
\caption{{\bf Density contrast.} Equatorial projection of the density contrast $\delta$ at initial time $\tls$ (left panel) and at present cosmic time $t_0$ (right panel). The vertical and horizontal scales correspond to $[x,y]=[R\cos\phi,\,R\sin\phi]$, with $R=a\,r$ (notice that $a=1$ at $t=\tls$). Solid radial line segments are the curves $\B_{+}(r)$. Dashed line segments mark the boundaries of the angular partitions $\Delta\phi_{ik}$. Dashed circles denote the radial comoving values $\chi_*^i$ that mark the comoving homogeneity spheres. Notice that in most of the volume the density contrast is roughly the background value $\delta\approx 0$ ($\rho \approx \bar{\rho}$) .}
\label{Fig2}
\end{figure*}
\subsection{Angular location of the spatial density maxima}\label{angular}   

Configurations containing $n$ density maxima (one in each interval $\Delta_*^i$ as in figure 8 of \cite{nuevo}) follow if we assign to each maxima the angular coordinates $(\theta_i,\phi_i)$ by prescribing at each $\Delta_*^i$ the dipole parameters in a piecewise manner:
$X=-\cos\phi_i{\sin\theta_i}\,\zeta_i\,f_i,\,\,Y=-\sin\phi_i{\sin\theta_i}\,\zeta_i\,f_i,\,\,Z=-\cos\theta_i\,\zeta_i\,f_i$ 
with the constants $0<\zeta_i<1$ controlling the density contrast amplitude. The $n$ functions $f_i(\chi)$ are thus given by
\begin{equation} f_i(\chi) = \sin^2\left[\frac{(\chi-\chi_*^{i-1})\pi}{\chi_*^i-\chi_*^{i-1}}\right],\label{fdef}\end{equation}
and satisfy the following boundary and smoothness conditions: $f_i(\chi_*^{i-1})=f_i(\chi_*^i)=0,\,f'_i(\chi_*^{i-1})=f'_i(\chi_*^i)=0$ and $f''_i(r_*^{i-1})=f''_i(r_*^i)=0$. Configurations that are more general than those examined in \cite{nuevo}, admitting several maxima in assorted angles $(\theta_{ik},\phi_{ik})$ within every $\Delta_*^i$, follow by defining $X,\,Y,\,Z$ (at each $\Delta_*^i$) as piecewise functions (see Table 1):   
$X=-\cos\phi_{ik}{\sin\theta_{ik}}\,\zeta_{ik}\,f_i,\,\,Y=-\sin\phi_{ik}{\sin\theta_{ik}}\,\zeta_{ik}\,f_i,\,\,Z=-\cos\theta_{ik}\,\zeta_{ik}\,f_i$ 
on a partition $\Delta\phi_{ik}=\psi_{i\,k}<\phi_{ik}<\psi_{i\,k+1}$ (with $k=1,..,p$) of angular domains (at each $\Delta_*^i$) separated by fixed azimuthal angles $\psi_{i\,k}$ whose value is chosen to fulfil smoothness conditions between each angular domain
\footnote{The angular partition at each shell $\Delta_*^i$ is equivalent to  matching several regions of separate Szekeres models with different dipole parameters along common surfaces marked by constant $\phi$, hence parametrised by $(t,r,\theta)$. Under certain algebraic restrictions on  $\zeta_{ik},\phi_{ik},\theta_{ik}$ the metric and its derivatives tangent to the matching surfaces are continuous, hence this matching can be smooth (Darmois conditions hold) even if the derivatives with respect to $\phi$ are discontinuous. For the case $Z=0$ considered in the numerical example Darmois conditions hold without further restrictions.}.
The radial coordinate location is the same for all maxima in the same interval $\Delta_*^i$ and the constants $0<\zeta_{ik}<1$ define the density contrast amplitude of the maxima.  

For illustrative purposes we select $Z=0$, so that (from (\ref{sol1})--(\ref{sol2})) we have $\theta_{ik}=\pi/2$ and thus all spatial density maxima (whose existence is guaranteed by the choice of $\muqls$ and $\kaqls$) are located in the equatorial plane $\theta_\pm=\pi/2$ (the more general case $Z\ne 0$ is analogous). We select the parameters $X$ and $Y$ in the piecewise manner explained above and shown explicitly in Table 1. Since $f_4(\chi_*^4)=f'_4(\chi_*^4)=0$, we can choose $X=Y=0$ (and thus $\bW=0$) for $\chi>\chi_*^4$, from this radius the configuration becomes spherically symmetric and convergent at all $t$ to a $\Lambda$CDM background \cite{nuevo}. 
\begin{figure*}
\centering
\includegraphics[scale=0.59]{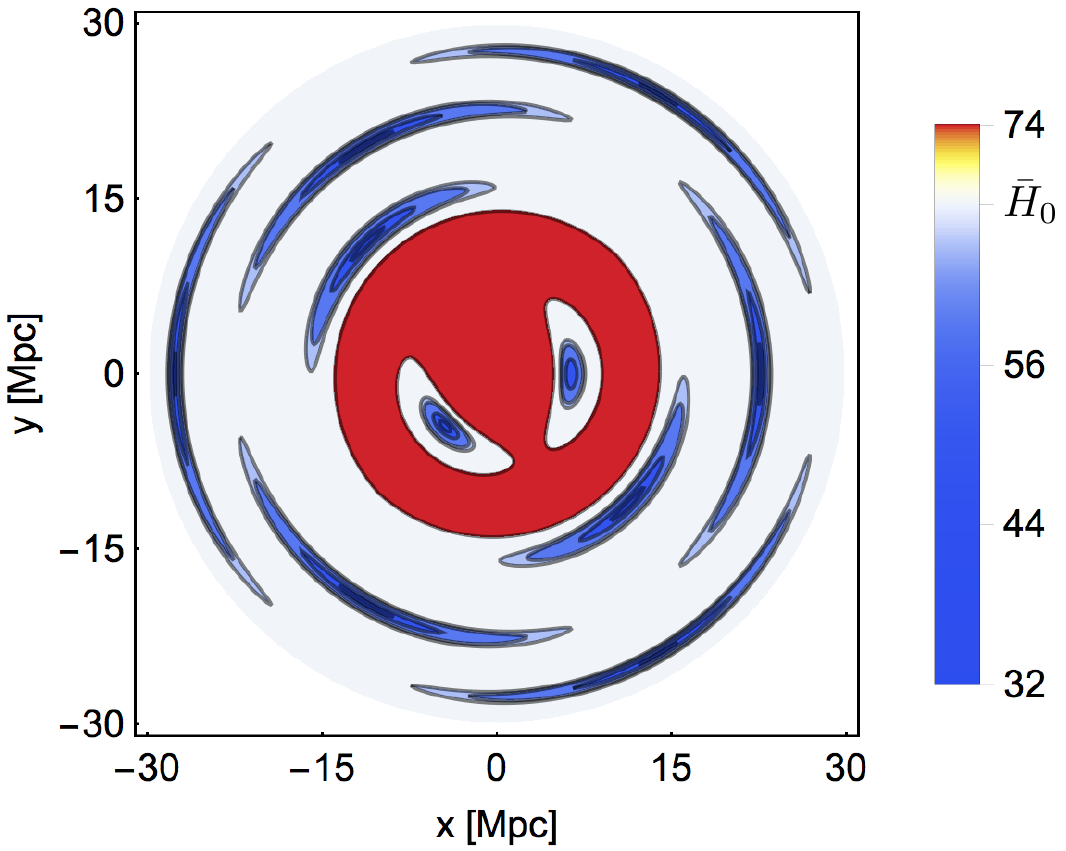}
\quad
\includegraphics[scale=0.40]{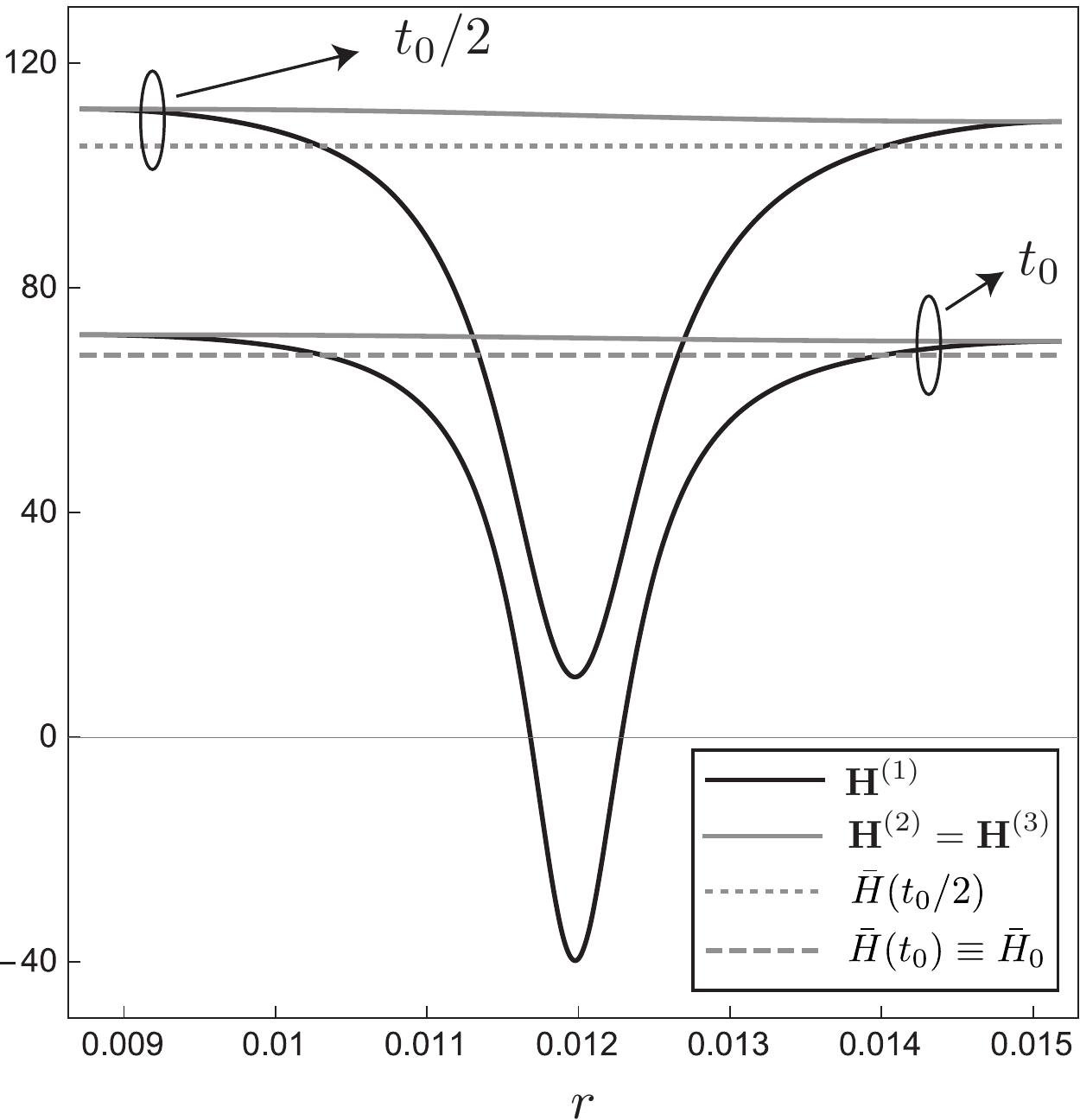}
\caption{{\bf The expansion scalar and the eigenvalues of the expansion tensor.} The figure depicts the equatorial projection of the Hubble scalar $\HH$ (left panel) evaluated at $t_0$ (in units km/(s Mpc)) and the eigenvalues $\Huno,\,\Hdos=\Htres$ of the expansion tensor (right panel) given by (\ref{haches}) evaluated along a radial ray (curve with $t,\theta,\phi$ constant), for $t=t_0/2,t_0$ and with $\theta=\theta_{21}=\pi/2,\phi=\phi_{21}=3\pi/4$ marking the angular coordinates of one of the over--densities listed in see Table 1 (similar curves result for all over--densities). Notice that $\HH$ as well as $\Hdos=\Htres$ are everywhere positive, taking (as expected) larger values in the void region than in the over--densities. However, $\Huno$ becomes negative at $t=t_0$ along the over--densities, thus indicating that the latter are undergoing a ``pancake'' collapse.}
\end{figure*}
\begin{figure*}
\centering
\includegraphics[scale=0.55]{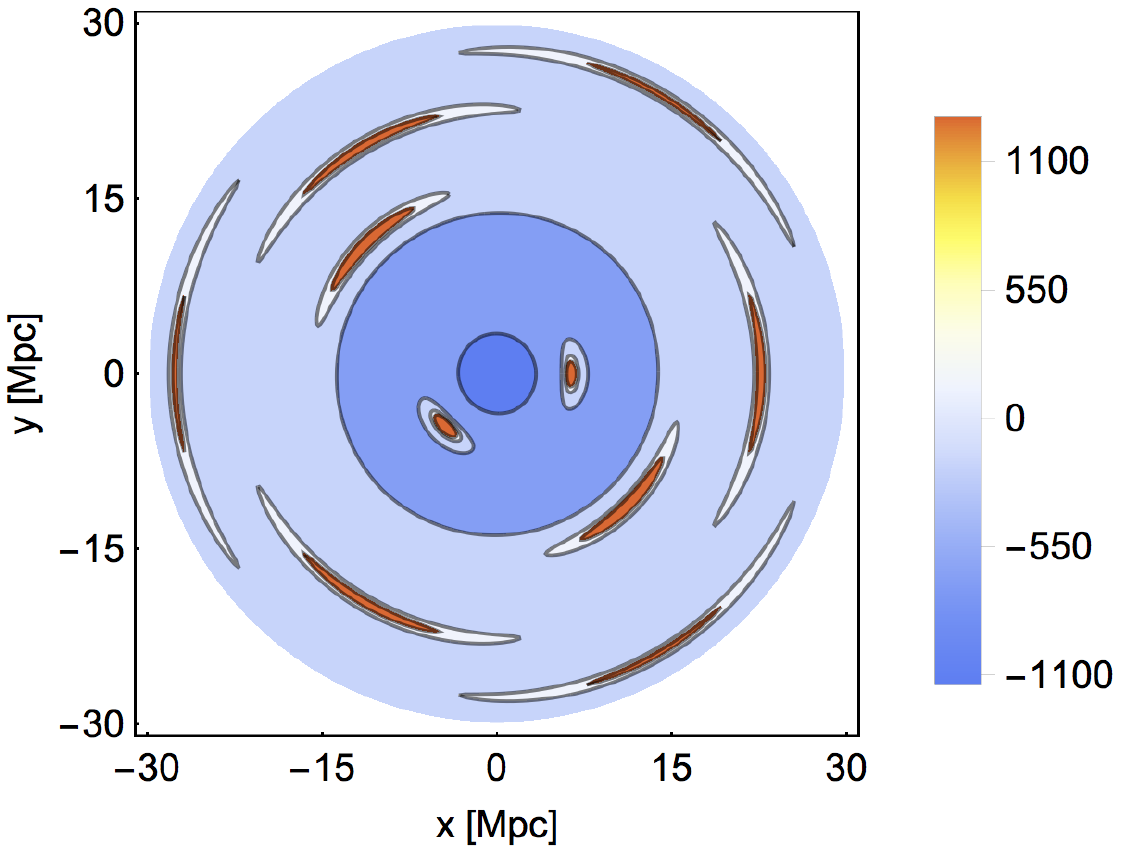}
\quad
\includegraphics[scale=0.55]{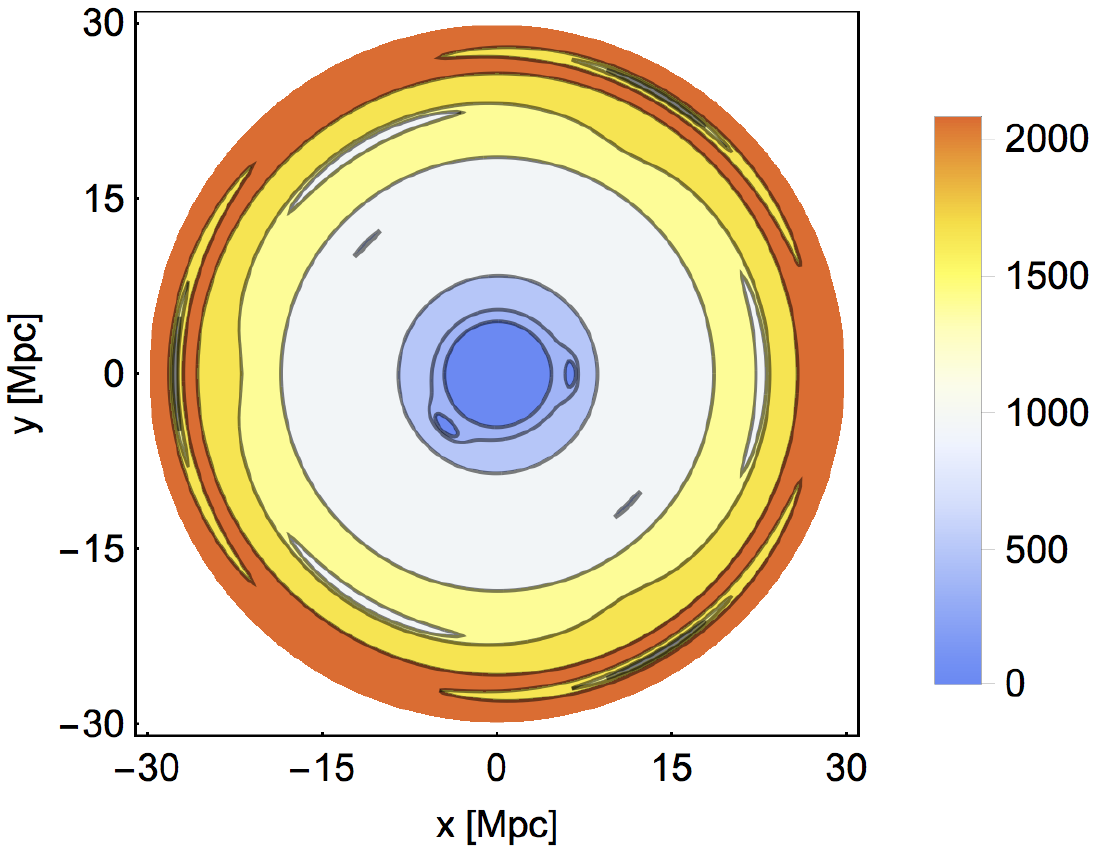}
\caption{{\bf Peculiar velocities.} Equatorial projection of the radial peculiar velocity field (in km/s) at the present cosmic time $t_0$, with respect to the  background identified as the CMB frame (left panel) and with respect to the observer at the centre of the void (right panel). }
\label{Fig4}
\end{figure*}

\section{Discussion.}\label{discussion}

By integrating the system (\ref{FFq1})-(\ref{constraints2}) for the initial conditions specified in Table 1, we can examine relevant dynamical quantities that characterise these multi--structure configurations. It is important to mention that these configurations are not spherically symmetric, hence the apparent rotational symmetry in the resulting graphics displayed in figures 2, 3 and 4 is merely an effect arising from employing spherical coordinates. This effect disappears when the figures are plotted in terms of proper radial distance (see figure 1 of \cite{nuevo}) or luminosity distance. 

\subsection{Density contrast.}\label{contrast}

We obtain the density contrast $\delta=(\mu-\bar\mu)/\bar\mu$, where $\mu=(4\pi\rho)/(3\bar\HH^2)$ and $\bar\mu(t)=(4\pi\bar\rho)/(3\bar\HH^2)=\bar\Omega^m(t)/2$, with the Szekeres density obtained from $\rho=\rho_q[1+\Drho]$ and $\bar\rho(t),\,\bar\HH(t)$ are the density and Hubble scalar of the $\Lambda$CDM background. Figure 2 displays the level curves of the equatorial projection of $\delta$, as functions of the area distance $R=a(t,r)r$, at the initial time $t=\tls$ (left panel) and at present time $t_0=13.7$ Gys (right panel). Both panels reveal in most of the spherical volume a slight under--density $\delta<0$ with near background density ($\delta\approx 0$), together with well defined and localised structures: a spheroidal density void around the origin (blue shading) surrounded by ten elongated over--densities (red/yellow shading), each one around a local density maximum located in the azimuthal angles given in Table 1, in the curves $\B_{+}(r)$ (solid line segments) in each one of the angular partitions in each of the four intervals $\Delta_*^i$. As shown in \cite{nuevo}, the over--densities have a pancake shape 3--dimensional morphology. Notice that the initial multi--structure shape is preserved in time, but it expanded from $\sim$0.08\,Mpc at $\tls$ to about  $\sim$ 80 Mpc at $t_0$, the negative amplitude of the density contrast of the void increased three orders of magnitude from $\delta_{\textrm{\tiny{LS}}}\sim -2\times 10^{-4}$ at $\tls$ to $\delta_0\sim -0.4$ at $t=t_0$, while the density contrast amplitudes of the over--densities evolved from the linear value $\delta_{\textrm{\tiny{LS}}}\sim 3.7\times 10^{-4}$ to fully non--linear values $\delta_0\sim 1.5-3$, as expected for $\sim 10-20$\, Mpc sized structures.

\subsection{Expansion and collapse of Szekeres structures.}\label{collapse}

The criterion for local collapse in inhomogeneous models follows from the signs of the eigenvalues $\Hii,\,((A)=1,2,3)$ of the expansion tensor $\HH^a_b=\HH\,h^a_b+\sigma^a_b$ (see equations (5)--(6) of \cite{pantano}). Since the shear tensor is traceless, the Hubble scalar $\HH=(1/3)(\Huno+\Hdos+\Htres)$ is the simple arithmetic average expansion. The isotropic spheroidal collapse (three negative eigenvalues) implies $\HH<0$, but ``pancake'' collapse (one negative eigenvalue) and ``filamentary'' collapse (two negative eigenvalues) can occur with $\HH>0$ (overall average expansion). 

For Szekeres models the eigenvalues of $\HH^a_b$ take the form  
\footnote{The eigenvalues of $\HH^a_b,\,\sigma^a_b$ and $h^a_b$ were computed for the metric (\ref{g1})--(\ref{g3}). Since they are coordinate independent invariant quantities, they are independent of the choice of metric components.}
\begin{equation}  \Huno = \HH+2\DDh=\HH_q +3\DDh,\qquad \Hdos=\Htres = \HH-\DDh=\HH_q,\label{haches}\end{equation}
where we remind the reader that $\HH_q=\dot a/a$ and $\DDh=(1/3)\dot\GG/\GG$ follow directly from (\ref{FFq5})--(\ref{FFq6}).  It is straightforward to obtain the eigenvalues $\Hii$ given by (\ref{haches}), evaluated at $t=t_0$, from the numerical solution of the system (\ref{FFq1})--(\ref{constraints2}). We the depict in Figure 3 the equatorial projection of the Hubble scalar $\HH$ plotted in terms of the area distance $R_0=a_0\,r$ (left panel), as well as the eigenvalues $\Huno_0$ and $\Hdos_0=\Htres_0$ evaluated for fixed $t=t_0/2$ and $t=t_0$ along a radial ray intersecting one of the over--densities of figure 2. All these quantities are given in units of $\hbox{km}/(\hbox{s\,Mpc})$. The left panel of figure 3 reveals that most of the displayed volume expands at a slightly larger but almost background value $\HH_0\approx \bar H_0 = 68\, \hbox{km}/(\hbox{s\,Mpc})$. Both panels  also reveal a strong anti--correlation between the Hubble flow associated with $\Hii_0$ and $\HH_0$ and the density field of figure 2: 
\begin{itemize}
\item the maximum of $\HH_0$ takes the value of $\sim 73\, \hbox{km}/(\hbox{s\,Mpc})$, roughly in the same location as the density minimum in the void centre, denoting the fastest expansion rate in the central void. The maxima of $\Hii_0$ (not displayed) occur also in the central void and have similar magnitudes. 
\item  the minima of $\HH_0$ in the left panel of figure 3 roughly coincide with the density maxima in the right panel of figure 2, in agreement with the expected slower expansion rate of expanding over--densities. Regarding the expansion eigenvalues at the over--density (right panel), the minima of $\Hdos_0=\Htres_0$ remain positive with values close to the background $\HH_0\sim 60\,\hbox{km}/(\hbox{s\,Mpc})$, but the minimum of $\Huno_0$  becomes negative ($\sim -40\,\hbox{km}/(\hbox{s\,Mpc})$). In fact, $\Huno_0<0$ holds for all the over--densities, thus indicating unequivocally that these structures have started undergoing a pancake collapse at $t=t_0$. Further, as shown in figure 5, these structures end up collapsing into a pancake shaped shell crossing singularity at times much later than $t_0$. However, the virialisation process occurs before these singularities are approached, indicating how the  description of structure formation by means of Szekeres models breaks down (as with the spherical collapse model).    
\end{itemize}   
The inhomogeneous Hubble scalar in the left panel of figure 3 reveals an average deviation of about $\sim \pm 10-20\,\%$ from the background CMB based value $\bar H_0\sim 68\,\hbox{km}/(\hbox{s\,Mpc})$, which is compatible with the results of \cite{HubbleFlow1} in which inhomogeneities were modelled by Newtonian numerical simulations.    

\subsection{Radial peculiar velocities.}\label{peculiar}

The radial peculiar velocities of the structures relative to the background Hubble flow (identified with the CMB frame) can be computed from $\vpcmb=[\HH_0\,a_0-\bar\HH_0\bar a_0](\chi-\chi_b)$, where $\chi_b\gg \chi_*^4$ is a sufficiently large value of the normalised comoving radius $\chi$ that can be identified with the asymptotic $\Lambda$CDM background, so that $\vpcmb(\chi_b)\approx 0$. These velocities are depicted in the left panel of figure 4, showing low density regions expanding away from the background frame at $\vpcmb\sim -$\,1200\,km/s, while the over--densities fall into this frame at $\sim\,$1000--1200 \,km/s.  The latter are not comparable to our CMB dipole velocity $\sim 370$\,km/s, as they are infall velocities of $10-20$\,Mpc structures into the $\Lambda$CDM background and thus do not take into account infall velocities of observers inside these structures with respect to their centres of mass. Instead, the peculiar velocities of the over--densities in figure 4 should be compared with the estimated infall velocity $\vpcmb\sim$\,600 km/s of our local group with respect to the $\Lambda$CDM background. In fact, these velocities are roughly compatible with similar velocities reported for the range of length scales under consideration: from data and observations \cite{pecvel1,pecvel2} and from numerical simulations \cite{pecvel31,pecvel32,pecvel33}.

Radial peculiar velocities with respect to an observer at the void centre comoving with the origin (depicted by the right panel) are computed from $\vpvoid=(\HH-\HH|_{r=0})\,a_0r$. These velocities exhibit an expansion away that is roughly linearly proportional to the radial area distance to the void centre, reaching $\vpvoid\sim$\,2000\,km/s for structures located $\sim$\,30\,Mpc away. As shown in the right panel of figure 4, these velocities closely match the peculiar velocities observed \cite{pecvel4} for galaxies in the Virgo supercluster with respect to an observer in the centre of the local void.   
\begin{figure}
\centering
\includegraphics[scale=0.4]{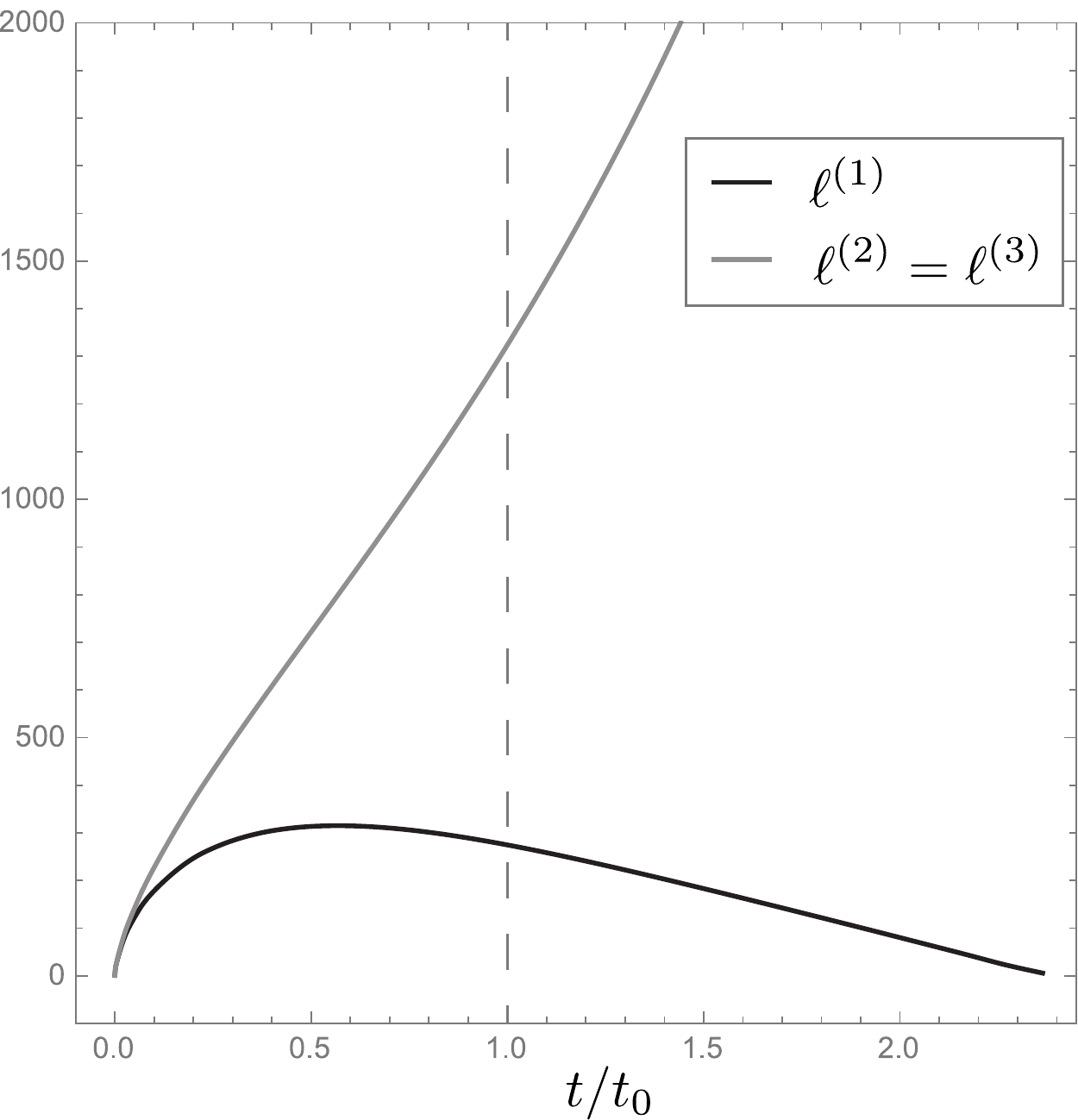}
\caption{{\bf Scale factors and pancake collapse.} The figure displays the scale factors given by (\ref{ells}), plotted as as functions of $t/t_0$ for the spatial coordinates of the over--density of the right panel of figure 3. Notice the pancake collapse of the structure in which the scale factor $\ell^{(1)}$ (associated with $\Huno$) is already decreasing at $t/t_0=1$ , while $\ell^{(2)}=\ell^{(3)}$  (associated with $\Hdos=\Htres$) are increasing for all $t$. Eventually a shell crossing singularity occurs as $\ell^{(1)}\to 0$ at some $t=t_{\rm{\tiny{sx}}} \sim 3t_0$, when the Szekeres description of structure formation breaks down.}
\label{Fig5}
\end{figure}

\section{Theoretical issues}

\subsection{An exact relativistic analogue of the Zeldovich approximation.}

The pancake collapse of the Szekeres structures suggests a non--trivial connection to the Newtonian Zeldovich Approximation (ZA) \cite{ZA1,ZA2}. Since the Szekeres models are an exact solution of Einstein's equations, this issue has been examined in various attempts to obtain relativistic generalisation of the ZA \cite{Zeld11,Zeld12,Zeld13,Zeld21,Zeld22,Zeld23}. We provide here a brief qualitative discussion that reveals how the Szekeres pancake collapse yields an exact relativistic analogue of the ZA  (for comprehensive treatment see \cite{Zeld11,Zeld12,Zeld13,Zeld21,Zeld22,Zeld23}). 

In Newtonian gravity the ``displacement'' relating Eulerian $y^n, n=1,2,3$ and Lagrangian coordinates $x^n$ for a homogeneously expanding medium is $y^n=\bar a(t) x^n$
\footnote{These coordinates bear no relation with the $[x,y]$ coordinates used in figures 2, 3 and 4.}
. The ZA considers for dust sources a first order correction through a  displacement $y^n=\bar a(t)[x^n+\Psi^n(t,x_m)]$ (with $\Psi(t_\ini,x^m)=0$). Assuming $\Psi^n_{,m}$ to be symmetric we obtain the following form of the density \cite{ZA1,ZA2}
\begin{equation} \rho = \frac{\rho_\ini}{\hbox{det}\,y^n_{,m}}=\frac{\rho_\ini}{\bar a^3[1-\xi^{(1)}][1-\xi^{(2)}][1-\xi^{(2)}]},\label{rhoZA}\end{equation}
where $-\xi^{(A)}(t,x^m)$ ($A=1,2,3)$ are the eigenvalues of the ``deformation'' tensor $\xi^m_n=\Psi^m_{,n}$ whose components take a simple form in the coordinates $x^A$ that diagonalise $\Psi^n_{,m}$. Since for arbitrary displacements $\Psi^n$ the three eigenvalues are different (say $0<\xi^{(3)}<\xi^{(2)}<\xi^{(1)}$), eventually as they grow in time we have $\xi^{(1)}\to 1$ while $\xi^{(2)},\,\xi^{(3)}<1$, producing a ``wall'' or ``pancake'' shape deformation in which distances contract in the direction of the eigenvector of $\xi^{(1)}$ and expand in the directions of the other two eigenvectors, leading in the end to a density caustic or singularity ($\rho\to\infty$ as $\det\,y^n_{,m}\to 0$). 

A direct qualitative comparison of the above mentioned process can be established with the evolution of Szekeres models through the exact density form
\begin{equation}\rho=\frac{\rho_\ini}{\ell^{(1)}\ell^{(2)}\ell^{(3)}} =\rho_i\,\frac{\JJ_\ini}{\JJ},\qquad \JJ=\sqrt{\hbox{det}\,g_{mn}}=\frac{a^3\,(\Gamma-\bW)\,r\,\sin\theta}{\sqrt{1-\KK_{qi}r^2}},\label{rhoSZ}\end{equation}
where $g_{mn}$ is the spatial metric in (\ref{g1})--(\ref{g2}) and the normalised scale factors $\ell^{(A)}$ follow from the eigenvalues $\Hii=\dot\ell^{(A)}/\ell^{(A)}$ of the expansion tensor $\HH^a_b$ in (\ref{haches}) with the condition $\ell_\ini^{(A)}=1$  
\begin{equation}\ell^{(1)} = a\,\GG=\frac{a\,[\Gamma-\bW]}{1-\bW},\qquad \ell^{(2)}=\ell^{(3)}=a.\label{ells}\end{equation}
Comparison of (\ref{rhoZA}) and (\ref{rhoSZ}) yields immediately 
\begin{equation} \xi^{(1)} = 1-\frac{\ell^{(1)}}{\bar a}=1-\frac{a}{\bar a} \frac{\Gamma-\bW}{1-\bW},\qquad  \xi^{(2)} = \xi^{(3)} = 1-\frac{\ell^{(2)}}{\bar a}=1-\frac{a}{\bar a}, \end{equation}
so that the 3--dimensional deformation matrix takes the from $\xi^A_B = \xi^{(1)}\delta^A_1\delta^1_B+\xi^{(2)}\left[\delta^A_2\delta^2_B+\delta^A_3\delta^3_B\right]$, where $x^A$ are the coordinates that diagonalise the spatial metric (see \cite{sussbol} and Appendix A of \cite{nuevo}). The Szekeres configurations we have studied provide an exact relativistic analogue of the ZA, as the pancake collapse of the structures occurs in an analogue manner as described before: at $t=t_\ini$ we have zero deformation $\xi^{(A)}=0$, but as shown in figure 5, we have along the over--densities $\ell^{(2)}>\ell^{(1)}$ as these scale factors grow for $t>t_\ini$, so that $\ell^{(1)}\to 0$ and $\rho\to\infty$ (shell crossing) may occur as $\Gamma-\bW\to 0$ at some $t_{\rm{{sx}}}>t_0>t_\ini$ with $\ell_{\rm{{sx}}}^{(2)}=\ell^{(3)}_{\rm{{sx}}}=a(t_{\rm{{sx}}},r)> 0$. Hence, the over--densities we have studied exhibit the same type of pancake deformation and collapse as Newtonian structures studied by means of the ZA, leading also to a final singularity (shell crossing) as $\xi^{(1)}\to 1$ occurs as $t\to t_{\rm{{sx}}}$, while $0<\xi_{\rm{{sx}}}^{(2)}=\xi_{\rm{{sx}}}^{(3)}<1$ holds. 

\subsection{Connection with linear perturbations.}

When considering the early evolution of inhomogeneities, the deviation from an FLRW background is small and we may present the inhomogeneities  as perturbations of the otherwise $\Lambda$CDM homogeneous universe, itself described by the background quantities $\bar{A}(t)= \{\bar\rho, \bar\HH, \bar\KK\}$. The perturbative description is valid for a regime where, given a small positive parameter $\epsilon \ll 1$, the following relations hold at a given initial time time $t_\ini=\tls$:
\begin{equation}
\label{smallness:1}
| A_{q} (r,\tls) - \bar{A}(\tls) | \approx \mathcal{O}(\epsilon), \qquad
 | r A_{q} '(r,\tls) | \approx \mathcal{O}(\epsilon).
\end{equation}
Since at the initial time we assume $\als =\Gls =  1$, the above conditions imposed on Eqs.~\eqref{DDa} and \eqref{Drho} imply,
\begin{equation}
\label{smallness:2}
| \DDals(r,\theta,\phi)| \approx \epsilon, \qquad  | \Drhols(r,\theta,\phi)| \approx \mathcal{\epsilon}\quad\Rightarrow\quad |\Als(r,\theta,\phi)-\barAls|\approx \mathcal{\epsilon}, 
\end{equation}
Further, it can be shown that for cosmic time intervals sufficiently close to $\tls$ the metric variables $a$ and $\Gamma$ will satisfy for all $r$ that (see proof in \cite{perts})
\begin{equation}
\label{smallness:3}
a - \bar{a}(t) \approx \mathcal{O}(\epsilon), \qquad \Gamma - 1 \approx \mathcal{O}(\epsilon).
\end{equation}
As a consequence of \eqref{smallness:1}--\eqref{smallness:3}, up to first order in $\epsilon$ the evolution equations \eqref{FFq1} and \eqref{FFq2}  are identical to the energy conservation and Raychaudhuri equations of the $\Lambda$CDM background, while the constraint \eqref{constraints1} is the background Friedman equation and \eqref{FFq5}--\eqref{FFq6} are the definitions of $\bar\HH$ and $\DDh$. The remaining evolution equations \eqref{FFq3} and \eqref{FFq4} and the constraint \eqref{constraints2} can then be linearised up to first order in  $\epsilon$, leading to:
\begin{eqnarray}
\dot\Delta^{(\rho)} &=& -3\,\DDh + \mathcal{O}(\epsilon^2)\label{FFq3lin} \\
 \dot {\textrm{\bf{D}}}^{(\HH)} &=&  -2 \bar \HH\, \DDh-\frac{4\pi}{3}\bar\rho\,\Drho + \mathcal{O}(\epsilon^2),\label{FFq4lin}\\
 \DDKK &=& \frac{8\pi}{3}\bar\rho\Drho-2\bar \HH \DDh + \mathcal{O}(\epsilon^2),\label{cons2lin}
\end{eqnarray}
which combine into the second order equation:
\begin{equation} \ddot\Delta^{(\rho)}+2\bar\HH\,\dot\Delta^{(\rho)}-4\pi\bar\rho\Drho + \mathcal{O}(\epsilon^2)=0.\label{Delta2}\end{equation}
Evidently, \eqref{FFq3lin}--\eqref{Delta2} are mathematically equivalent to the linear dynamical equations of the Cosmological Perturbation Theory (CPT) formalism in the isochronous gauge (cf. \cite{Mukhanov:1990me,malik:wands}). However, the exact fluctuations $\Drho$ and $\DDa$ relate to the gradients of $A_q$ via \eqref{DDa}--\eqref{Drho}, and thus are analogous but not strictly equivalent to CPT perturbations. The rigorous equivalence between Szekeres scalars and CPT variables follows from extending to Szekeres the results obtained in \cite{perts} for LTB models. Instead of the q--scalars $A_q$ in (\ref{qscals}), we consider  the  functional averages $\Aav_q[r_b]$ in bounded comoving domains $\DD[r_b]$ along the time slices, which define the non--local exact fluctuations 
\begin{equation}\Drhonl[r_b] =\frac{\rho(t,x^j) - \rhoav_q[r_b](t) }{ \rhoav_q[r_b]},\qquad \DDanl=A(t,x^j)-\Aav[r_b](t),\quad A=\rho,\,\HH,\,\KK,\label{nonlocfluc}\end{equation}
For models admitting an asymptotic $\Lambda$CDM background, the scalars of the CPT background become rigorously defined by the functional averages $\Aav_q[r_b]$ evaluated in asymptotic domains extending to the complete time slices through the asymptotic limit 
\begin{equation}\lim_{r_b\to\infty}\left\{\rhoav_q[r_b],\HHav[r_b],\KKav [r_b] \right\}=\left\{\rhobaras,\HHbaras,\KKbaras\right\}=\left\{\bar\rho,\bar\HH,\bar\KK\right\}.\end{equation}
while the non-local exact fluctuations in \eqref{nonlocfluc} lead in this limit to the asymptotic exact fluctuations:
\begin{equation}
 \Drhoas=\frac{\rho(t,x^j)-\rhobaras(t)}{\rhobaras(t)}=\lim_{r_b\to\infty}\Drhonl[r_b],\qquad \DDaas=A(t,x^j)-\Abaras(t)=\lim_{r_b\to\infty}\DDanl[r_b]
\end{equation}
so that $\Drhoas=\delta=\rho/\bar\rho-1$ (but $\ne \Drho$), where $\delta$ is the density contrast plotted in figure 2 and obtained from the exact Szekeres density $1+\delta=[\rho_q/\rhobaras]\,[1+\Drho]$. Actually, the conditions in eqs. \eqref{smallness:2} and \eqref{smallness:3} guarantee that the difference is of order $\epsilon$ and consequently, their amplitude too:
\begin{equation}
\Drhoas \approx \Drho, \qquad 
\Dhas \approx \DDh, \qquad 
\DKKas \approx \DDKK.
\end{equation}
The evolution equations for these variables are mathematically identical to \eqref{FFq3lin}--\eqref{Delta2} up to order $\epsilon$ (see \cite{perts}). As a consequence, the evolution of the linear order quantities is exactly the well known evolution of the dust perturbations in the synchronous and comoving gauge (see e.g. \cite{Sussman:2013qya}). 

The equivalences presented in this subsection show unequivocally that the exact inhomogeneities of the Szekeres models can be directly connected with the perturbative approach to the study of large scale structure formation.         

\section{Conclusions.} 

We have shown how the dynamical freedom of Szekeres models makes it possible to obtain a fully relativistic non--perturbative description of non--trivial networks of cold dark matter structures (over--densities and density voids) evolving from the last scattering surface to the present. In particular, we provided a numerical example of the evolution of a $\sim$ 80\,Mpc sized region immersed in a $\Lambda$CDM background, consisting of a spheroidal density void surrounded by ten pancake shaped density maxima, placed at given radial and angular comoving locations, specified by the initial conditions displayed in Table 1. This configuration (whose density contrast is depicted in figure 2) represents a huge improvement over previous attempts to model cosmic structure with LTB models \cite{kras2,BKHC2009,BolHel} or Szkeres models of class I \cite{Bsz1,Bsz2,BoSu2011} and class II \cite{meures}, as they furnish a significantly better (though still coarse grained) description of cosmic structures observed, or inferred, at a $\sim$100 Mpc scale today (as for example in \cite{cosmography1,cosmography2}).

By looking at the Hubble scalar $\HH$ and the eigenvalues of the expansion tensor $\Huno,\,\Hdos=\Htres$, we have shown through a numerical example how an $\sim$ 80 Mpc region that expands on average ($\HH>0$ for all $t$) contains various $\sim$ 10--20 Mpc over--densities undergoing a local ``pancake'' collapse at present cosmic time ($\Huno_0<0,\,\Hdos_0=\Htres_0>0$). We have also examined for these structures the radial peculiar velocities with respect to the CMB background frame and with respect to an observer in the void centre. These velocities fit very well observed velocities reported in the literature for same scale structures.

We have also shown how the resulting Szekeres models relate to other standard theoretical frameworks considered in current cosmological research. Specifically, we show (i) how the pancake collapse of the Szekeres over--densities provides an exact relativistic analogue to the pancake collapse in the Newtonian Zeldovich approximation, and (ii) how their evolution in the linear regime relates to cosmological  perturbations of dust sources in the synchronous gauge. 

As mentioned before (last paragraph of section \ref{networks}), our models are not related to Swiss Cheese models. Rather, they generalise the ``walls and voids'' models examined in \cite{mattsson} and the ``onion'' models of \cite{onion}, all based on LTB solutions (though the onion model has been applied to simple dipole Szekeres models in \cite{kokhan,mock}). Since the scalar averaging of non--spherical Szekeres scalars is spherically symmetric \cite{sussbol} (in Buchert's formalism and in quasi--local averaging), then the scalar averaging of the structures we have studied should yield similar results as the averaged LTB walls and voids models of \cite{mattsson}. However, this issue needs to be carefully verified and thus will be examined in a separate article.        

While the possibility of describing the relativistic and non--perturbative evolution of elaborated networks of non--spherical over--densities and voids is very appealing, the Szekeres models we have studied exhibit the expected  limitations characteristic of all analytic or semi--analytic structure formation models. Evidently, it is wholly unreasonable to expect these models to describe the virialisation of cosmic structures or complex dynamical interactions like the ``bullet cluster'' or mergers of structures. As the over--densities undergo a pancake collapse the models break down when the shell crossing singularity is approached, and thus the description of the virialisation process must be introduced ``by hand'' as is done with the spherical collapse model. Still, notwithstanding the above mentioned limitations,  these models have an enormous potential for application in open problems of current cosmological research, such as:   
\begin{enumerate}
\item Exploring the effects of relativistic corrections in cosmographic studies \cite{cosmography1,cosmography2} and Newtonian simulations (e.g. \cite{N-body1}), as well as the effects of non--linearity in perturbative relativistic treatments \cite{Adamek:2014xba,Adamek:2013wja,Fidler:2015npa}.
\item Addressing the apparent tension between estimations of the Hubble constant from the CMB and from supernovae surveys \cite{H0tension}, and the associated problem of the differential expansion produced by nearby nonlinear structures \cite{dflow11,dflow12,dflow2}. 
\item Verifying if Szekeres models allow for the description of collapse regimes besides pancake collapse (spherical or filamentary). Explore in more detail the connection between evolving Szekeres structures and fully relativistic generalisations of the Zeldovich approximation \cite{relpecvel21,relpecvel22,Zeld11,Zeld12,Zeld13,Zeld21,Zeld22,Zeld23} and implementing them \cite{Adamek:2014xba,Adamek:2013wja,Fidler:2015npa} in specific structure formation scenarios.
\item A nonlinear relativistic treatment of peculiar velocities \cite{relpecvel1,relpecvel21,relpecvel22} and interpretation of the observed redshift space distortions \cite{zdistort1,zdistort2}. This is absent in the literature and the Szekeres models can help to fill this gap too. 
\item Exploring the relativistic corrections in structure formation scenarios examined by means of Newtonian simulations \cite{Grcorr1,Grcorr21,Grcorr22} and those attributed to modified gravity \cite{ModGrav1,ModGrav2}, as well as the correspondence and equivalence of exact solutions vs linear and non--linear perturbative approaches \cite{Dictionary1,Dictionary2} or an extension of previous work on these issues in LTB models \cite{perts}.
\item  The Szekeres models we have examined can serve as non--trivial exact ``test models'' to probe important issues theoretical, such as back--reaction, averaging and the ``fitting problem''  \cite{BR1,BR2,BR3,BR4}.  Besides looking at a more realistic non--spherical extension of the results of \cite{mattsson}, we can use the Szekeres configurations as ``test models'' to examine controversial theoretical issues on back--reaction \cite{GW1,GW2,GW3,antiGW}.     
\end{enumerate} 
These possible applications will be pursued in separate articles currently under  elaboration for future submission.  



\acknowledgments
The authors acknowledge support from research grants PAPIIT--UNAM IA101414 and IA103616, as well as SEP-CONACYT 239639. This work was undertaken entirely at ICF-UNAM. I.D.G. acknowledges support from CONACYT program of doctoral grants.


\end{document}